\documentclass[conference]{IEEEtran}

\IEEEoverridecommandlockouts
\usepackage{verbatim}
\usepackage{longtable}
\usepackage{adjustbox}
\usepackage{amsmath}
\usepackage{amsfonts}
\usepackage{amssymb}
\usepackage{multirow}
\usepackage[T1]{fontenc}
\usepackage{amsmath,amssymb,amsfonts}
\usepackage{fancyhdr}
\usepackage[ruled,vlined,linesnumbered]{algorithm2e}
\usepackage{graphicx}
\usepackage{textcomp}
\usepackage{xcolor}
\usepackage{subcaption}
\usepackage{algpseudocode}
\usepackage{enumitem}
\usepackage{float}
\usepackage{graphicx}
\usepackage[normalem]{ulem}
\usepackage{url}
\useunder{\uline}{\ul}{}
\begin{document}

\title{FlexiContracts: A Novel and Efficient Scheme for Upgrading Smart Contracts in Ethereum Blockchain\\
}

\author{\IEEEauthorblockN{Tahrim Hossain$^{1}$, Sakib Hassan$^{2}$, Faisal Haque Bappy$^{3}$, Muhammad Nur Yanhaona$^{4}$,\\ Sarker Ahmed Rumee$^{5}$, Moinul Zaber$^{6}$, and Tariqul Islam$^{7}$}
\IEEEauthorblockA{
$^{1, 3, 7}$ Syracuse University, USA\\
$ ^{2, 5, 6}$ University Of Dhaka, Bangladesh\\
$ ^{4}$ BRAC University, Bangladesh\\
$ ^{6}$ United Nations University, Portugal\\
Email: mhossa22@syr.edu, skbhssn.sh@gmail.com, fbappy@syr.edu, nur.yanhaona@bracu.ac.bd, \\rumee@cse.du.ac.bd, zaber@unu.edu, and mtislam@syr.edu} 
}

% \author{
% \IEEEauthorblockN{Tahrim Hossain}
% \IEEEauthorblockA{
% \textit{Syracuse University}\\
% Syracuse, New York, United States \\
% mhossa22@syr.edu}
% \and
% \IEEEauthorblockN{Sakib Hassan}
% \IEEEauthorblockA{
% \textit{University Of Dhaka}\\
% Dhaka, Bangladesh \\
% skbhssn.sh@gmail.com}
% \and
% \IEEEauthorblockN{Faisal Haque Bappy}
% \IEEEauthorblockA{
% \textit{Syracuse University}\\
% Syracuse, New York, United States \\
% fbappy@syr.edu}
% \and
% \IEEEauthorblockN{Muhammad Nur Yanhaona}
% \IEEEauthorblockA{
% \textit{BRAC University}\\
% Dhaka,Bangladesh \\
% nur.yanhaona@bracu.ac.bd
% }
% \and
% \IEEEauthorblockN{Tariqul Islam}
% \IEEEauthorblockA{
% \textit{Syracuse University}\\
% Syracuse, New York, United States \\
% mtislam@syr.edu}
% }
\maketitle

\thispagestyle{fancy}
\lhead{This work has been accepted at the IEEE International Conference on Trust, Security and Privacy in Computing and Communications (TrustCom 2024)}

\begin{abstract}
Blockchain technology has revolutionized contractual processes, enhancing efficiency and trust through smart contracts. Ethereum, as a pioneer in this domain, offers a platform for decentralized applications but is challenged by the immutability of smart contracts, which makes upgrades cumbersome. Existing design patterns, while addressing upgradability, introduce complexity, increased development effort, and higher gas costs, thus limiting their effectiveness. In response, we introduce \textit{FlexiContracts}, an innovative scheme that reimagines the evolution of smart contracts on Ethereum. By enabling secure, in-place upgrades without losing historical data, \textit{FlexiContracts} surpasses existing approaches, introducing a previously unexplored path in smart contract evolution. Its streamlined design transcends the limitations of current design patterns by simplifying smart contract development, eliminating the need for extensive upfront planning, and significantly reducing the complexity of the design process. This advancement fosters an environment for continuous improvement and adaptation to new requirements, redefining the possibilities for dynamic, upgradable smart contracts. 
\end{abstract}


\section{Introduction}
In recent times, the advent of Blockchain technology has significantly transformed numerous industries by offering secure and transparent decentralized systems. At the forefront of this technological evolution is Ethereum \cite{buterin2013ethereum}, a blockchain platform that facilitates the creation of decentralized applications (DApps) through the use of smart contracts. Smart contracts are self-executing agreements encoded with predefined rules and conditions, enabling trustless interactions between parties \cite{szabo1996smart}. The transformative impact of blockchain on real world application areas such as finance, supply chain management, and healthcare \cite{tapscott2016Blockchain, lim2021literature, kuo2017blockchain} demonstrates its potential to redefine traditional business models and operational processes.

Smart contracts predate blockchain, with initial challenges of lacking a trusted execution platform and reliable result sharing \cite{szabo1996smart}. Ethereum resolved these by facilitating decentralized code execution on its global blockchain \cite{buterin2013ethereum}. Smart contract immutability, while ensuring security, poses challenges when updates are necessary \cite{atzei2017survey}. Studies highlight how this rigidity impedes modifications for evolving business needs, creating a gap between technological promise and practical usability \cite{Nzuva2019SmartCI,Khan2021}. Adapting to new business deals or legal requirements often requires costly reviews \cite{MikSmartContract}. This immutability proves restrictive in dynamic contexts and hinders post-deployment bug fixes. These issues have prompted research into smart contract upgradability, aiming to balance security with flexibility.

The blockchain community has faced major issues from smart contract bugs. Notable incidents include ``The DAO" reentrancy attack, which led to a loss of over 50 million USD in Ether \cite{daoxploit2016}, and the parity wallet attack, where an access control error trapped about 500,000 Ether \cite{paritywalletmultisighack2017}. Additionally, a study by Torres et al. revealed that over 42,000 contracts suffer from integer overflow bugs, issues arising when arithmetic operations exceed the integer type's capacity,  impacting many ERC-20 token contracts \cite{osirisIntegerBugs}. These vulnerabilities have led to significant financial losses in tokens and ether and have prompted the community to focus on strengthening smart contract security through secure and transparent upgrade processes.

Developers currently use various design patterns to address the issue of smart contract immutability \cite{santipall,meisami2023comprehensive}. Examples of such patterns include the proxy pattern \cite{openzeppelin_proxy_patterns}, the eternal storage pattern \cite{openzeppelin_eternal_storage}, metamorphic smart contracts \cite{frowis2022not}, and the diamond pattern \cite{diamondPattern}. However, these approaches introduce challenges like complex data migration, increased gas costs, centralization risks due to single-entity control, and complications in managing multiple contracts. Additionally, there's a risk of losing historical data during migration, affecting the contract's transparency and trustworthiness.

The need for mutable smart contracts arises not only from the necessity of fixing bugs or adding new features but also from the imperative to adapt to evolving business conditions driven by socioeconomic, environmental, and political changes. This paper presents \textit{FlexiContracts}, designed to address the challenges of smart contract immutability, including fixing bugs and integrating new functionalities. \textit{FlexiContracts} aim to provide a secure upgrade process that simplifies design and deployment, maintains trust and transparency, and actively involves stakeholders in decision-making. 

The following are the contributions of our paper.

\textbf{Decentralized Governance Protocol.} We propose an on-chain governance protocol enabling decentralized stakeholder participation in smart contract upgrades via blockchain-recorded voting. 

\textbf{Automated Data Reorganization.} We propose a seamless storage reorganization mechanism that dynamically adjusts storage slots based on code change analysis, eliminating tedious data migration workflows and ensuring continuity across upgrades.

\textbf{Streamlined Upgrade Process.} We introduce a single-contract upgrade method that simplifies the development complexities by eliminating manual procedures and reducing data disruption.

We structure the remainder of this paper as follows: in Section 2, we critically review related work. In Section 3, we outline the system architecture of \textit{FlexiContracts}, covering theoretical and practical influences. In Section 4, we present a feature comparison to evaluate our approach against existing methods. We conclude the paper in Section 5 summarizing the key findings.

\section{Related Works}

Research on smart contract upgradability focuses on enabling upgrades without costly contract replacement or data migration, sparking debate over its merits. Critics argue that `upgradeability is a bug' \cite{upgradabilityIsABug}, while supporters believe it improves development, reduces costs, and allows for adaptability, making smart contracts more akin to traditional software lifecycles \cite{qasse2024immutable}. Previous research on smart contract upgradability has focused on off-chain upgrades \cite{fourTier}. Significant efforts, such as the development of Oyente by Luu et al. \cite{Luu}, have targeted pre-deployment vulnerability detection in smart contracts. Formal methods have also been applied for verification. Additionally, Rodler et al. introduced EVMPatch, an automated system that enables smart contract upgrades by rewriting bytecode \cite{evmpatch}. 

The Ethereum community uses several methods to allow smart contracts to be upgraded, employing different techniques and design patterns to help update deployed contracts. One common approach is to use proxy smart contracts \cite{openzeppelin_proxy_patterns}, which act as intermediaries that direct user requests to an upgradeable implementation contract. This setup allows the proxy’s address to remain unchanged, meaning users interact with the same proxy contract even when the underlying implementation is updated. This way, the user experience remains consistent while the functionality can improve without disruption. Another method is eternal storage \cite{openzeppelin_eternal_storage}, which separates data from logic, allowing for logic upgrades without affecting the stored data, thereby simplifying the upgrade process and enhancing safety. Additionally, metamorphic contracts utilize the \textbf{SELFDESTRUCT} command to terminate the old contract, release its resources, and deploy a new contract with updated logic at the same address. The diamond pattern \cite{diamondPattern} organizes functionalities into separate facets, each as an individual contract, all managed by a central diamond contract as the single entry point. This approach offers more flexibility and modularity than the proxy pattern, as it allows for independent updates to each facet. Overall, these various methods ensure that smart contracts on Ethereum can evolve and improve while maintaining a seamless experience for users.

Ethereum's default update method involves contract destruction and redeployment. Upgradability mechanisms add complexity due to additional code and architectural considerations, increasing bug risks and requiring extensive auditing \cite{salehi2022immutable,li2024characterizing,liu2024demystifying}. These approaches also increase gas costs, affecting usability and affordability of smart contract interactions \cite{salehi2022immutable,anto2023gas}. Many upgradability techniques rely on central authorities, introducing centralization risks and potential power abuse \cite{contractMigration}. Our methodology addresses these challenges by incorporating stakeholder participation through a voting mechanism and introducing seamless storage reorganization. This eliminates the need for complex, costly data migration, simplifies the smart contract upgrade and development process, and maintains minimal gas costs, all while preserving decentralization.

\begin{comment}
   \begin{figure}
   \centering
   %\includegraphics[width=1.0\linewidth]
   % \includegraphics[width=0.6\linewidth]
   \includegraphics[width=3.4in,height=1.55in]
   {figures/states.pdf}
   \caption{Proposal Lifecycle} % Caption for your figure
   \label{fig:proposal_lifecycle} % Assigning a label to the figure
\end{figure} 
\end{comment}

\begin{figure*}
    \centering
    \includegraphics[width=0.78\textwidth]{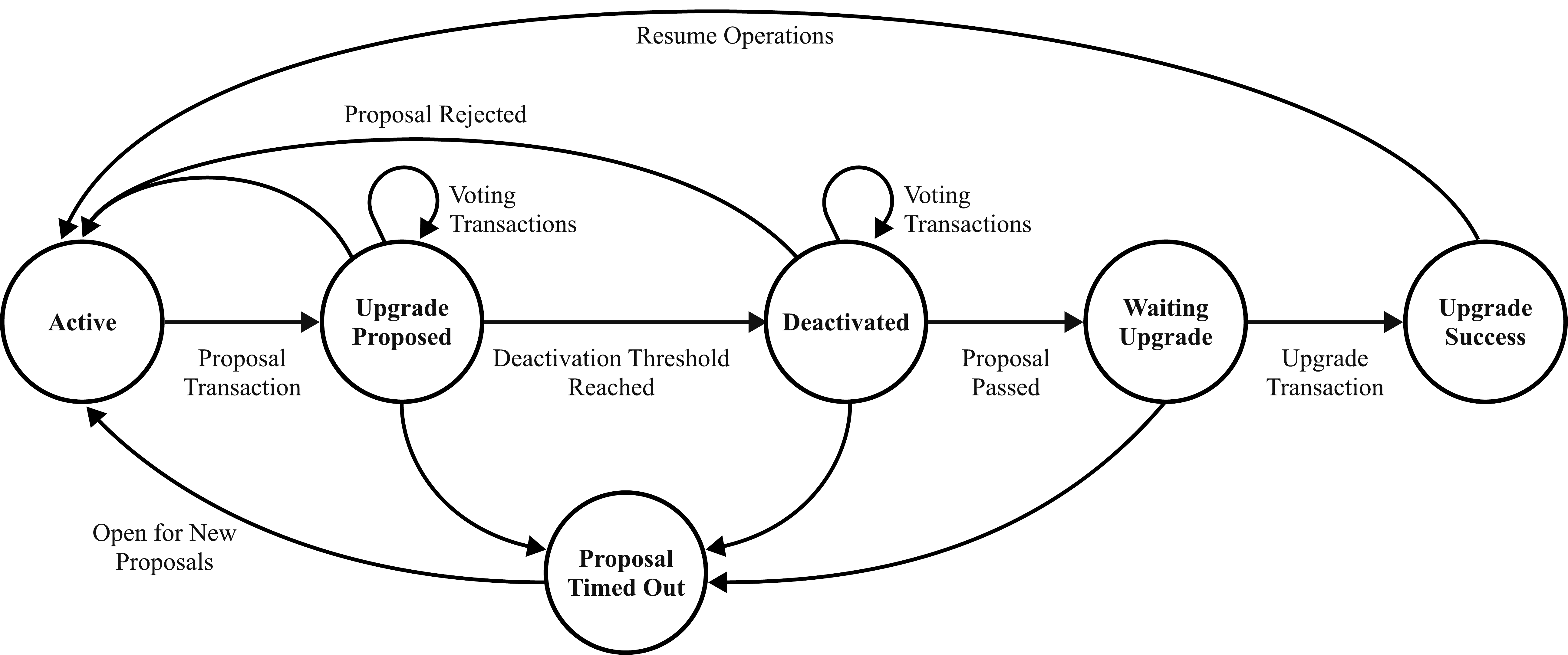}
    \caption{Proposal Lifecycle} 
    \label{fig:proposal_lifecycle} 
\end{figure*}

\section{System Architecture}
Our architecture for \textit{FlexiContracts} addresses the challenges of smart contract upgradability on Ethereum, where existing solutions often add complexity, increase gas costs, and pose security risks. To overcome these limitations, we combine a voting-based governance protocol with a strategic storage reorganization mechanism in \textit{FlexiContracts}. The governance protocol ensures secure and transparent stakeholder-driven decision-making. The storage reorganization mechanism ensures data integrity and continuity by automatically aligning existing data with upgraded contracts. We outline these components in this section, focusing on how they enable scalable, efficient, and secure contract upgrades.

\subsection{Voting-based Governance of Upgrade Proposals} 
We propose a voting-based governance protocol to manage smart contract upgrades effectively. This protocol involves enhancements to Ethereum's account structure to support mechanisms for tracking upgrade proposals, defining the criteria for proposal approval, and specifying the roles and permissions of stakeholders in the process. Additionally, our framework incorporates configurable parameters to manage the governance lifecycle, such as time constraints for resolving proposals and conditions under which the current contract version transitions to a non-executable state.

We manage the governance lifecycle using a systematic proposal workflow with a state-machine approach as illustrated in Fig. \ref{fig:proposal_lifecycle}. Once an upgrade is proposed, the smart contract enters a voting phase, exclusive to stakeholders. Following the voting process, the contract's state changes according to the result. Proposals that are rejected return to the previous version, whereas accepted ones transition to a non-executable status until the upgrade is finally implemented as depicted in Fig. \ref{fig:proposal_lifecycle}. Upon approval, a stakeholder initiates a final transaction, transitioning the contract to a state where the changes are applied, and the updated version of the contract becomes executable. This process ensures thorough deliberation and seamless integration of changes.

We strategically separate the initiation of smart contract upgrades from their eventual implementation, as illustrated in Fig. \ref{fig:proposal_lifecycle}. This creates distinct transactions for each phase. This design introduces a range of benefits. By separating the two processes, the architecture helps reduce uncertainty around resource use. This separation allows proposers to schedule upgrades during favorable network conditions, significantly reducing gas costs. It also removes the need for proposers to have significant account balances upfront, as the costs of proposing and implementing upgrades are spread out over different phases.

To reinforce the security of our governance framework, we design key parameters to enhance security and prevent misuse or exploitation of the proposal process. One such parameter is the time constraint, which enforces a predefined block-based timeframe for resolving proposals. This ensures that proposals do not remain unresolved indefinitely, reducing the risk of delays caused by inaction. Without this safeguard, adversarial stakeholders could exploit the system by repeatedly submitting proposals to deliberately delay critical upgrades. We specify another essential parameter that defines the conditions under which the current version of a contract transitions to a non-executable state. This is a crucial mechanism for addressing bugs, vulnerabilities, or other risks. This parameter allows stakeholders to halt a faulty version by achieving a predefined threshold of affirmative votes, preventing harmful operations while upgrades or fixes are applied. It ensures swift responses to critical issues while preventing misuse through significant stakeholder support, enhancing stability and confidence in governance.

Building on this foundation, our governance framework further empowers stakeholders by allowing dynamic refinement of these parameters. This adaptability ensures that the governance system remains responsive to evolving requirements and stakeholder needs. By fostering flexibility and security, it provides a robust foundation for managing smart contract upgrades effectively.

\begin{figure*}
    \centering
    \includegraphics[width=0.98\textwidth]{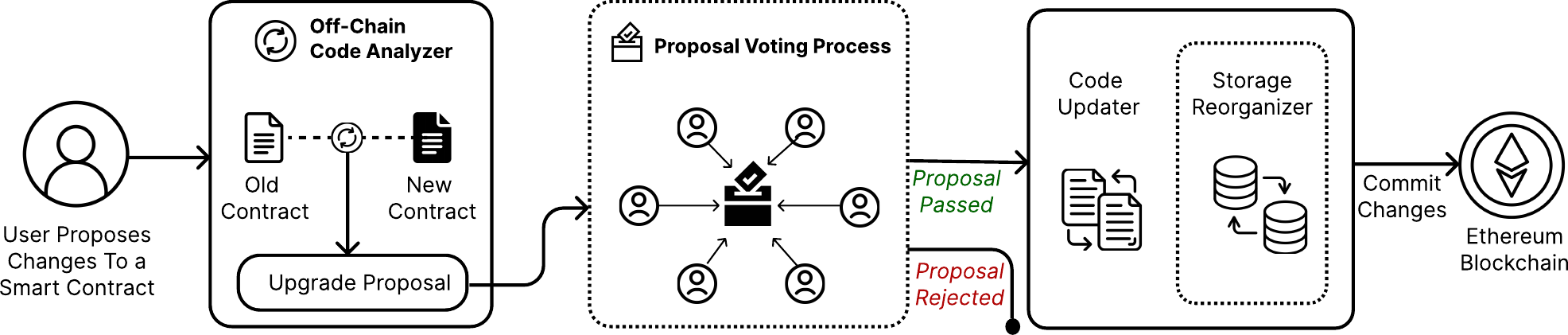}
    \caption{System Workflow} 
    \label{fig:workflow} 
\end{figure*}

\subsection{Strategic Data Reorganization for Upgrade Compatibility} 
\begin{figure}[h]
  \centering
  \begin{subfigure}[b]{0.48\columnwidth}
    \centering
    \includegraphics[width=0.65\textwidth]{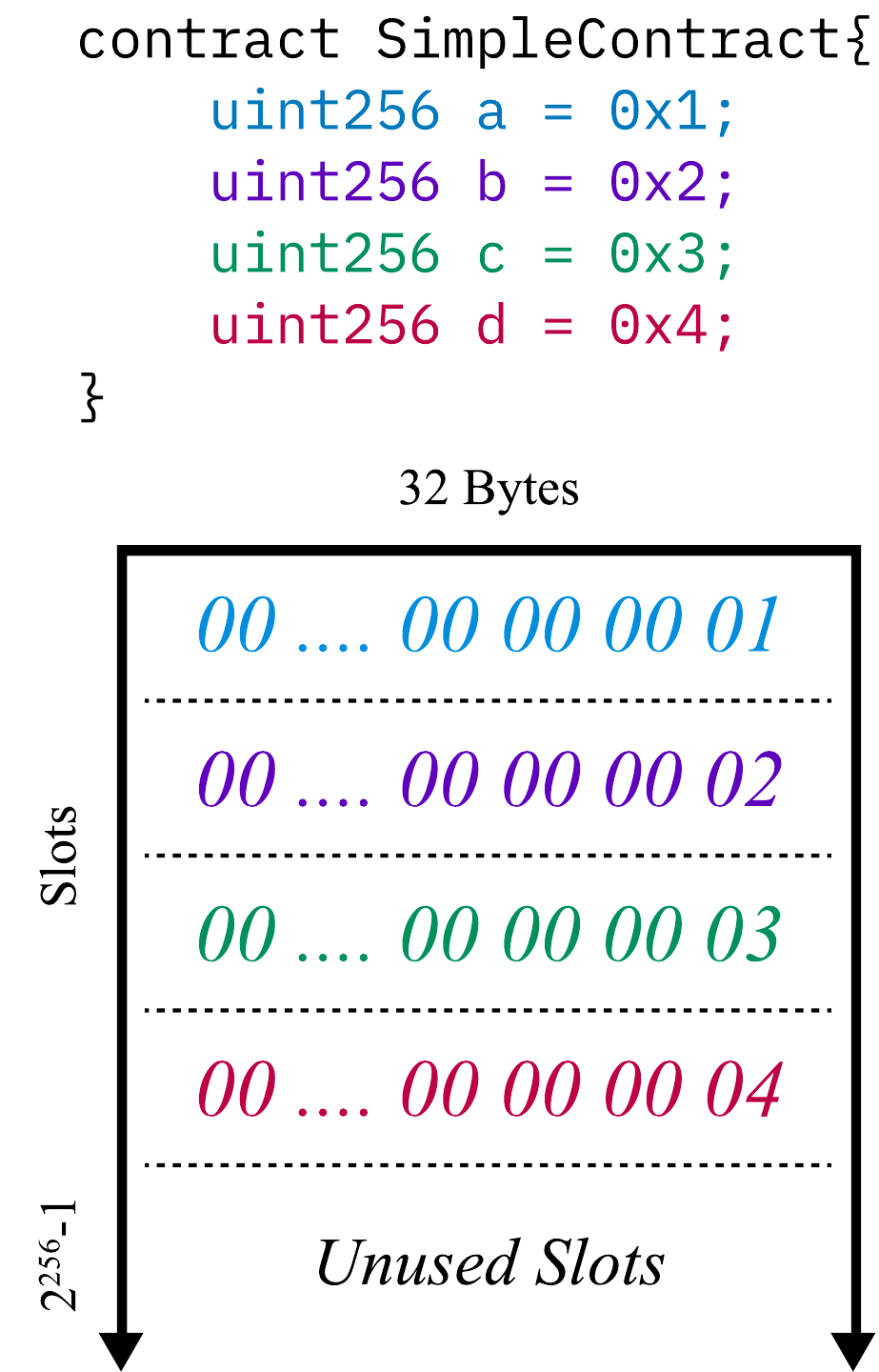}
    \caption{Storage Organization of Variables}
    \label{fig:storage_organization_variables}
  \end{subfigure}
  \hfill
  \begin{subfigure}[b]{0.48\columnwidth}
    \centering
    \includegraphics[width=0.9\textwidth]{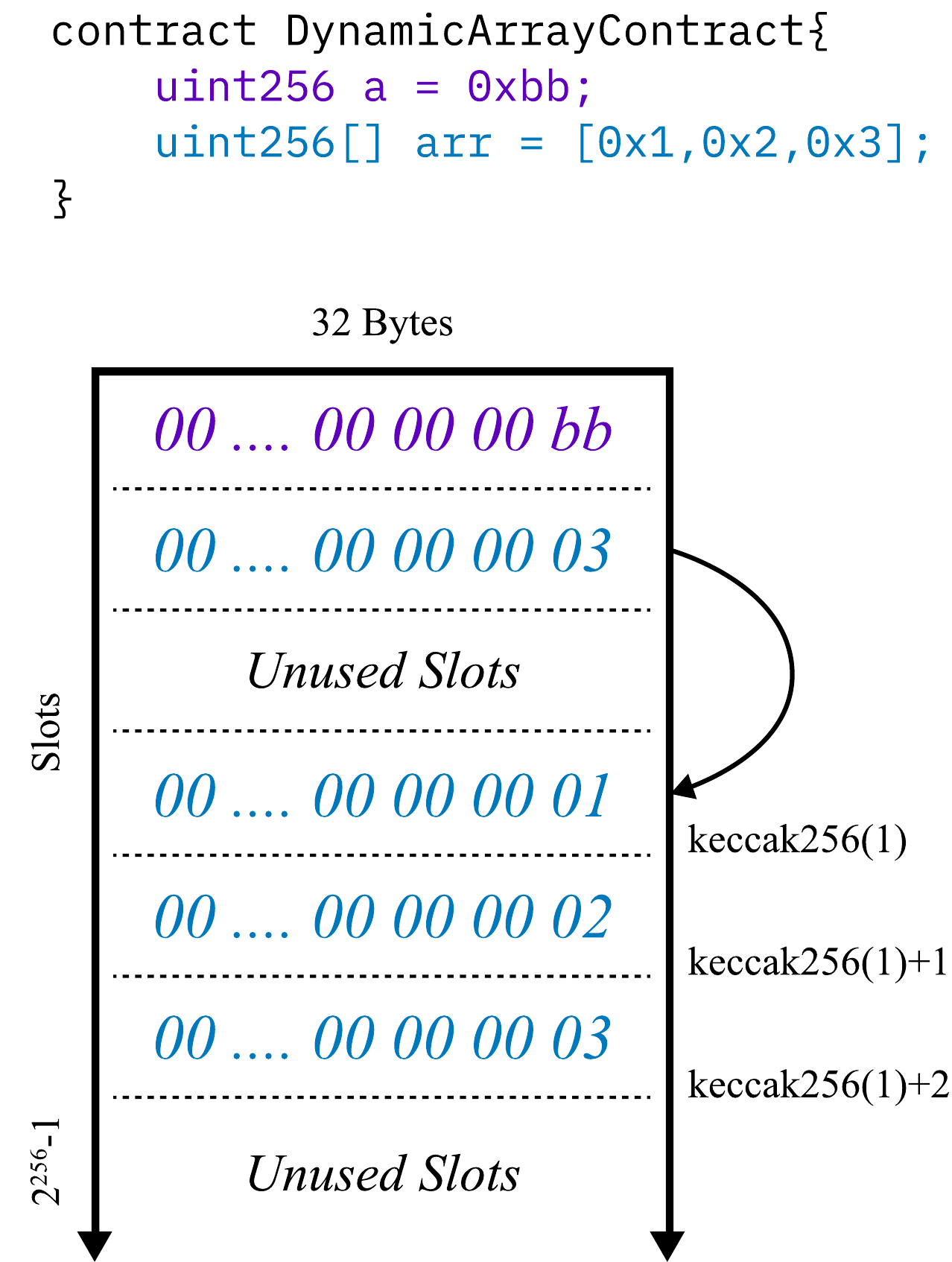}
    \caption{Storage Organization of Dynamic Array}
    \label{fig:storage_organization_dynamic_array}
  \end{subfigure}
  \label{fig:overallmemory}
  \caption{Storage Organizations}
\end{figure}

We propose a strategic data reorganization mechanism to ensure smart contracts can be upgraded without losing data integrity or operational continuity. By addressing the complexities of Ethereum's storage organization, this mechanism ensures seamless upgrades while maintaining consistency. Ethereum's storage consists of $2^{256}$ slots of 32 bytes each, managed by the Storage Trie for secure and efficient access. State variables in a smart contract are allocated storage slots systematically based on declaration order and size. Variables are tightly packed into 32-byte slots to optimize storage, while dynamic arrays and mappings follow specialized patterns for efficiency. In Fig. \ref{fig:storage_organization_variables} we demonstrate the sequential allocation of four 32-byte variables ($a$, $b$, $c$, $d$) to the first four slots based on their declaration order. Smaller types share slots while larger ones occupy their own. As an example, if four consecutive uint64 variables (8 bytes each) are defined, they would collectively occupy a single slot, as their combined size (32 bytes) matches the slot's capacity. 

Arrays and mappings employ sophisticated storage patterns for efficient data management. Fixed-size arrays store elements sequentially in contiguous slots starting from their declaration position, allowing elements smaller than 32 bytes to be packed together for storage efficiency. Dynamic arrays utilize a complex storage mechanism where the first slot, \textit{p} stores the array's length, and element positions are determined by hashing this slot's index \textit{keccak256(p)}, with elements stored from that computed location. In Fig. \ref{fig:storage_organization_dynamic_array} we demonstrate this storage protocol through an example of a dynamic array initialized with values \textit{[0x1, 0x2, 0x3]}. The storage layout allocates the array's length at its declaration position in slot with index 1, while the array elements are stored beginning at position \textit{keccak256(1)}, with subsequent elements occupying consecutive slots from this computed location. Mappings follow a similar hash-based approach, where each value's location is calculated using \textit{keccak256(k . p)}. Here, \textit{k} represents the key used to store a value in the mapping, and \textit{p} represents the storage slot index at which the mapping is declared. This ensures uniform distribution and efficient access while preventing direct iteration over values.

Given this intricate organization, any upgrade to a smart contract that alters the order or number of variables presents a significant challenge. Changes in the code could potentially misalign the data stored in the smart contracts's storage slots, leading to data corruption or loss. In \textit{FlexiContracts}, we address this by mapping old storage structures to new code requirements, ensuring accurate data placement post-upgrade while preserving data continuity. To achieve this, we incorporate two core components, an Off-Chain Code Analyzer and an On-Chain Storage Reorganizer, as illustrated in Fig. \ref{fig:workflow}.

\subsubsection{Off-Chain Code Analyzer}
We ensure a seamless transition from existing to proposed smart contract versions through this component, as shown in Fig. \ref{fig:workflow}. When a stakeholder submits a proposal to modify a smart contract, the analyzer examines the differences in storage structures between the current and proposed versions. It identifies modifications, additions, removals, or reordering of data elements within the smart contract's storage structure. Based on this analysis, it generates detailed instructions specifying the exact mapping of values from their current storage slots to the appropriate new storage slots. These instructions guide the on-chain storage reorganizer in modifying the contract’s storage layout to ensure compatibility with the new version. Once the instructions are generated, the proposal is forwarded for voting as depicted in Fig. \ref{fig:workflow}. This enables stakeholders to review and approve the changes.

\subsubsection{On-Chain Storage Reorganizer}
We simplify and automate the complex task of restructuring contract storage during upgrades through the on-chain storage reorganizer. Traditionally, upgrading a smart contract required deploying a new version and migrating data manually, often leading to errors. We eliminate this issue by directly reorganizing storage to align with the new contract through the on-chain storage reorganizer. This component operates alongside the Code Updater as shown in Fig. \ref{fig:workflow} and utilizes inputs from the off-chain code analyzer. The analyzer provides instructions specifying the required data movements and modifications, which the reorganizer executes to align the data with the new contract schema.

\begin{figure}[h]
   \centering
   \scalebox{0.7}{\includegraphics[width=0.45\textwidth]{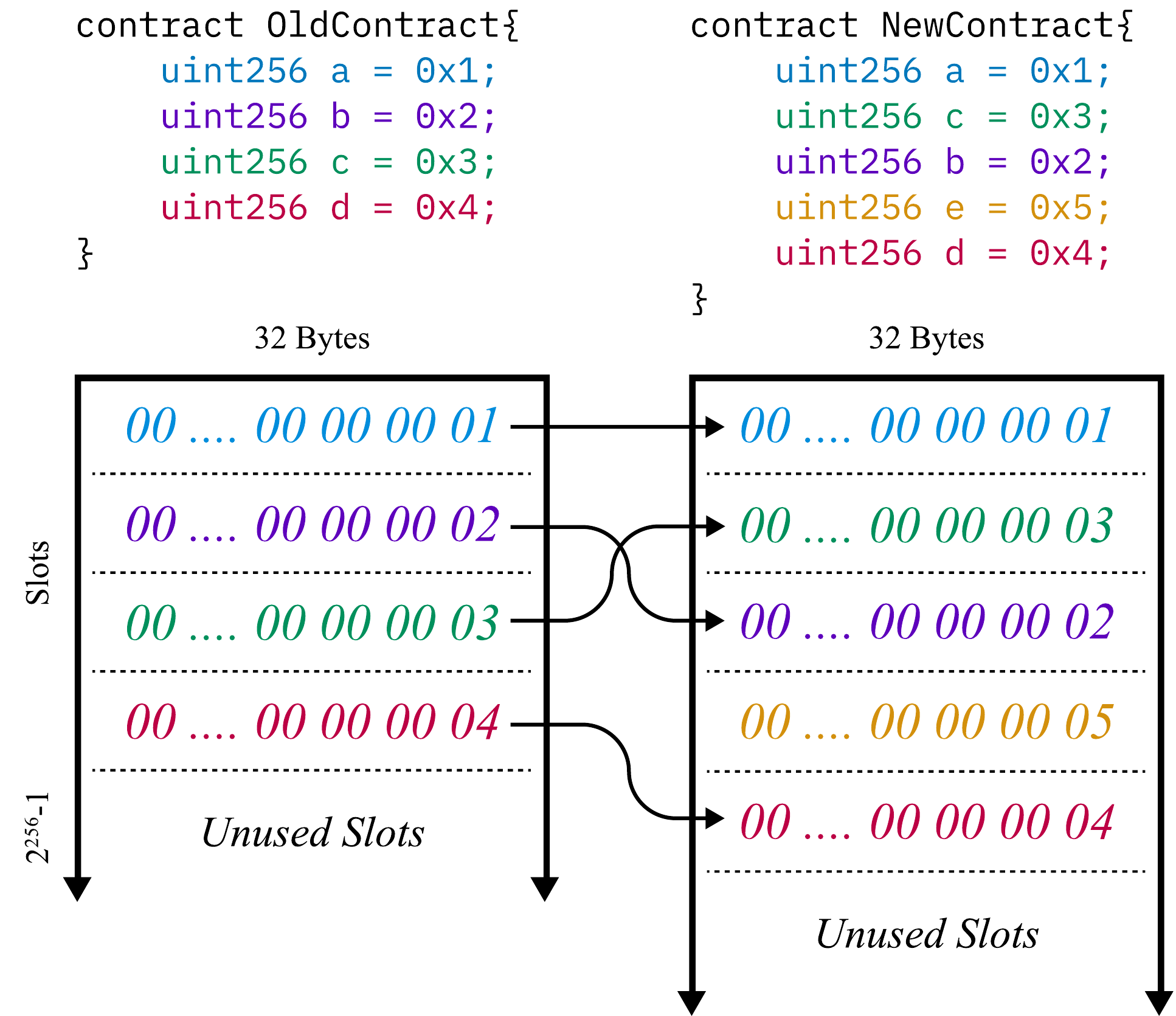}}
   \caption{Reorganization of value types} 
   \label{fig:storage_reorg_variables}
\end{figure}

To understand the role of the on-chain storage reorganizer in practice, it’s important to consider how different types of data stored in smart contracts require distinct approaches to reorganization. For instance, value types like integers or booleans may simply be reassigned to new storage slots to match the updated declaration order. In contrast, dynamic data structures such as arrays involve more intricate adjustments.

In Fig. \ref{fig:storage_reorg_variables} we illustrate how storage is reorganized during a smart contract upgrade. In the original contract, the variables \textit{a}, \textit{b}, \textit{c}, and \textit{d} are stored sequentially in slots 1 through 4, reflecting their declaration order. After the upgrade, we update the order using the reorganizer. \textit{a} remains in slot 1, \textit{c} moves to slot 2 taking \textit{b}’s original position. \textit{b} shifts to slot 3, which was previously occupied by \textit{c}. A new variable \textit{e} is introduced before \textit{d}, occupying slot 4 and pushing \textit{d} to slot 5.

\begin{figure}[h]
   \centering
   \scalebox{0.9}{\includegraphics[width=0.45\textwidth]{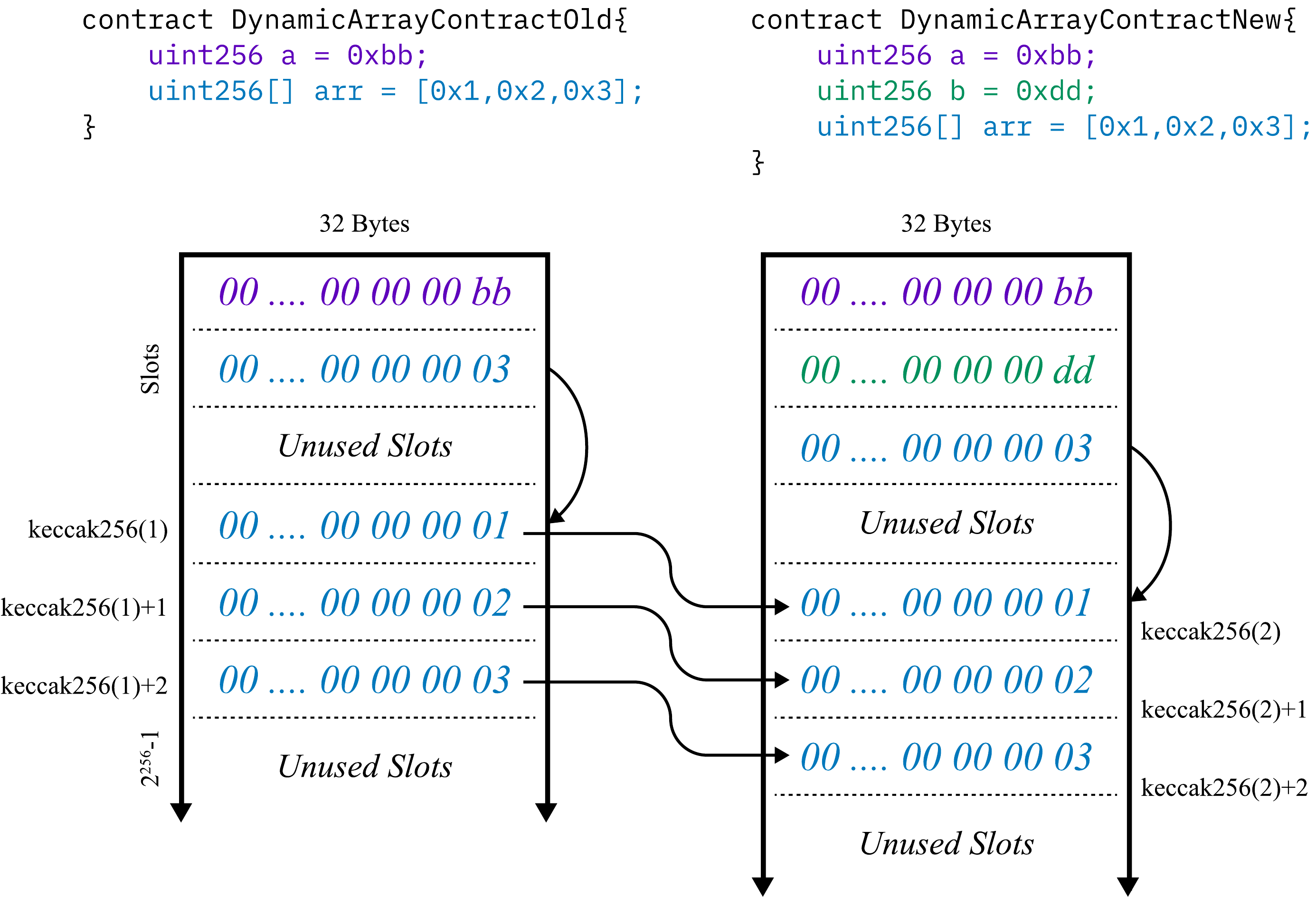}}
   \caption{Reorganization of dynamic array} 
   \label{fig:storage_reorg_dynamic_array}
\end{figure}

Reorganizing dynamic data structures like arrays require more complex adjustments. In Fig. \ref{fig:storage_reorg_dynamic_array} we show how arrays are reorganized during an upgrade. In the original contract, the size of the array is stored in slot 1, while its elements are stored sequentially starting at an address computed as \textit{keccak256(1)}, with subsequent elements located at \textit{keccak256(1)+1}, \textit{keccak256(1)+2}, and so on. When a new variable is introduced in the upgraded contract, the declaration order changes, shifting the base slot of the array from 1 to 2. This modification alters the starting address for the array's elements, which now begin at \textit{keccak256(2)}. To align the data with the new storage layout, we use the reorganizer to move the size metadata from slot 1 to slot 2 and recalculate the new positions for all array elements. It then transfers the elements to their corresponding slots starting at \textit{keccak256(2)}. This efficient handling of various data types demonstrates the capability of our on-chain storage reorganizer to ensure data integrity and consistency during upgrades, regardless of the complexity involved.

\section{Feature Comparison}
\textit{FlexiContracts} aims to streamline smart contract upgradability by addressing the limitations of existing patterns while maintaining security and efficiency. This section compares \textit{FlexiContracts} with established approaches, including proxy, eternal storage, metamorphic, and diamond patterns. By focusing on features such as state preservation, address continuity, and upgrade simplicity, we evaluate the strengths and trade-offs of these methods, highlighting the contributions of \textit{FlexiContracts} to the broader landscape of smart contract evolution.
\begin{table} [tbh]
\centering
\caption{Feature Comparison \textit{(FC = FlexiContracts, P = Proxy, ES = Eternal Storage, M = Metamorphic and D = Diamond)}}
\label{tab:example}
\resizebox{\linewidth}{!}{
\begin{tabular}{lccccc}
\hline
\multicolumn{1}{c}{\textbf{Criteria}}     & \textbf{FC}               & \textbf{P}                & \textbf{ES}               & \textbf{M}   &\textbf{D}             \\ \hline
Can replace entire logic         & $\checkmark$ & $\checkmark$ & $\checkmark$ & $\checkmark$ & $\checkmark$ \\ \hline
No need to migrate state from old contract  & $\checkmark$ & $\checkmark$ & $\checkmark$ & $\times$  & $\checkmark$   \\ \hline
User endpoint address unchanged             & $\checkmark$ & $\checkmark$ & $\times$     & $\checkmark$ & $\checkmark$ \\ \hline
No need to instrument source     & $\checkmark$ & $\times$     & $\times$     & $\times$ & $\times$     \\ \hline
No need to deploy a new contract to upgrade & $\checkmark$ & $\times$     & $\times$     & $\times$  & $\times$   \\ \hline
No indirection between contracts & $\checkmark$ & $\times$     & $\times$     & $\checkmark$ & $\times$ \\ \hline
\end{tabular}
}
\end{table}

An effective upgradability solution must address critical challenges without introducing unnecessary complexity or operational disruptions. In Table \ref{tab:example}, we highlight the critical aspects required for efficient upgradability of smart contracts. In-place logic changes eliminate the need for redeployment, ensuring smooth transitions without operational disruption. Address continuity ensures seamless integration with external systems and uninterrupted user interactions, preserving users' confidence in the integrity of the contract. The absence of complex instrumentation simplifies development by eliminating intricate setups, enabling developers to focus on core functionalities instead of upgrade complexities. This reduction in technical overhead minimizes the likelihood of errors during implementation, shortens development timelines, and decreases maintenance requirements, ultimately lowering both developer effort and operational costs. Finally, avoiding indirection between contracts reduces execution complexity by eliminating intermediary layers, lowering gas costs, and simplifying transaction paths.

In Table \ref{tab:example} we highlight key limitations in traditional upgradability methods. Proxy, eternal storage, and diamond patterns rely on indirection between contracts, increasing execution complexity and gas costs. These methods also require source instrumentation and the deployment of new contracts for upgrades, adding significant operational overhead. Metamorphic contracts, on the other hand, fail to preserve the state from old contracts, necessitating disruptive and error-prone data migration. Furthermore, eternal storage does not maintain the user endpoint address, leading to compatibility issues with external systems and user interfaces. \textit{FlexiContracts} addresses all these shortcomings, ensuring no state migration is required, preserving user endpoint addresses, and eliminating both source instrumentation and contract indirection. These features make \textit{FlexiContracts} a comprehensive solution for smart contract upgradability, offering streamlined and efficient upgrades without the trade-offs observed in traditional methods. 

\section{Conclusion} 
In this paper, we introduced \textit{FlexiContracts}, a novel approach to enhancing smart contract upgradability on Ethereum. By combining automated storage reorganization with on-chain governance, \textit{FlexiContracts} enables efficient and secure upgrades while simplifying the modification of deployed contracts. Its governance model ensures transparent, community-driven decision-making, and its automated storage reorganization preserves data continuity. In summary, \textit{FlexiContracts} provides a robust solution for advancing smart contract upgradability, optimizing efficiency, and enabling seamless integration of complex functionalities.

\bibliographystyle{IEEEtran}
\bibliography{IEEEabrv,references}
\end{document}